\documentclass[twocolumn,prd, floatfix]{revtex4}
\usepackage{epsfig,amsmath,amsthm,amssymb}
\newcommand{\Section} [1] {\section{#1} \setcounter{figure}{0} \setcounter{equation}{0}}

\begin{document}

\title{Neutrinos from Auger sources}

\author{Francis Halzen and Aongus \'O Murchadha}
\affiliation{Department of Physics, University of Wisconsin, Madison WI 53706}
\date{\today}

\begin{abstract}
The Pierre Auger observatory has presented evidence that the arrival directions of cosmic rays with energies in excess of $6\times10^{7}$ TeV may be correlated with nearby active galactic nuclei (AGN). In this context we revisit a suggestion based on gamma ray observations that nearby Fanaroff-Riley I galaxies such as Cen A and M87 are the sources of the local cosmic rays. We compute the accompanying neutrino flux and find a flux within reach of second-generation kilometer-scale neutrino telescopes.
 \end{abstract}
 
 \maketitle
 
\section{Gamma-ray Observations}

Recently the Pierre Auger observatory has presented evidence that the arrival directions of cosmic rays with energies in excess of $6\times10^{7}$ TeV may be correlated with nearby active galactic nuclei (AGN)\,\cite{auger}. The skymap suggest that events cluster in the direction of the Fanaroff-Riley I (FRI) radio galaxy Centaurus A. This source, along with the other nearby FRI M87, has previously been singled out as a potential cosmic ray accelerator on the basis of gamma-ray data\,\cite{anchordoquicena}. We revisit those arguments here as well as asking the question whether or not kilometer-scale neutrino observatories have the capability to unambiguously detect these sources.

It has been speculated for some time that supermassive black holes at the centers of active galaxies power particle flows that create the opportunity for particle acceleration to super-EeV energy\,\cite{ghs}. Both the accretion flow near the black hole nucleus and the extended jets are potential sites for shocks that allow for efficient particle acceleration. Blazars, a subclass of FRI whose jets are oriented along our line of sight, are abundant sources of TeV energy photons. Exceptionally, the nearby (16 Mpc) FRI M87 was observed in the late 1990's by the HEGRA stereoscopic system of five imaging atmospheric Cherenkov telescopes despite the fact that its jet is angled to more than $30^{\circ}$ from our line of sight\,\cite{Horiuchi}. The integrated flux observed is\,\cite{hegra}
\begin{displaymath}
F_{\gamma} (E_{\gamma} > 0.73\,{\rm TeV}) = 0.96 \pm 0.23 \times{\rm 10^{-12}\,cm^{-2}\,s^{-1}}
\end{displaymath}
Recently, the H.E.S.S. array obtained a limit of $10^{-12}$ in the same units on Cen A\,\cite{cenahess}, a similar source in the southern hemisphere at a distance of only 3.4 Mpc. Evidence for TeV emission from Cen A was obtained in the early 1970's with the Narrabri optical intensity interferometer of the University of Sydney\,\cite{cenasydney}, with an average flux of
\begin{displaymath}
F_{\gamma} (E_{\gamma}  > 0.3\,{\rm TeV}) = 4.4 \pm 1.0 \times{\rm 10^{-11}\,cm^{-2}\,s^{-1}}
\end{displaymath}
The variable flux was collected in two periods of heightened activity lasting roughly one year, pointing at a region of coherent emission of size of order 0.3 pc. This is consistent with the idea that the high energy emission is from an isotropic region near the central black hole of mass $M_{BH} = 2\times 10^{8}$ solar masses, about two orders of magnitude more massive than the one at the center of our galaxy\,\cite{anchordoquicena}. The fact that the detector beam did not include the radio lobes at the ends of the jets further supports the idea of a central engine at the base of the jets. The TeV fluxes of Cen A and M87 imply similar source luminosities of roughly $7\times10^{40}\, {\rm erg\,s^{-1}}$, assuming an $E^{-2}$ gamma-ray spectrum. This suggest that they are generic AGN, a fact we will exploit to construct the diffuse neutrino flux from all FRI.

\Section{Modeling the Sources}

We will assume that the photon flux is of pionic origin, i.e. that the high energy protons hinted at by the Auger data produce neutral pions on the gas surrounding the central black hole that decay to produce the detected high-energy photons. This is a plausible model given that the proton interaction length on a particle density of $ n\sim10^{6}\,{\rm cm^{-3}}$ is 4.6 pc, not much larger than the source of size $R\sim 0.3$\,pc. We conservatively assume a density similar to the one near the center of our own galaxy. The probability that the accelerated proton interacts is therefore expected to be of order $P\sim R/\lambda_{int}\sim R\,n\,\sigma_{prot}\sim 10\%$. The energy produced in a one year burst is
\begin{displaymath}
E_{\rm burst}=PGM_{BH}M_{\rm infall}/R.
\end{displaymath}
This leads to $M_{\rm infall}\sim 5\times10^{-5}$ solar masses. This is two orders of magnitude larger than the estimated infall at the center of our own galaxy\,\cite{Genzel} and is reasonable because of the burst nature and the larger mass of the Cen A black hole. We tentatively conclude that the central accelerator in FRI such as Cen A and M87 feeds a beam dump and pursue the implication of the Auger and TeV gamma ray observations for the emission of neutrinos, inevitably produced by the charged pions produced in association with the neutral ones\,\cite{agnnu}. Given that the black hole in Cen A is more massive than the one at the center of our own galaxy by two orders of magnitude, our assumption for the target density should therefore be considered a lower limit and values of $P$ larger by up to two orders of magnitude should be acceptable.

We will first estimate the neutrino flux from individual nearby sources and subsequently derive a diffuse flux assuming that the nearby sources are representative of the extragalactic source population. We finally comment on the suggestion that the pions are produced in photohadronic ($p\gamma$) rather than hadronic ($pp$) interactions\,\cite{hannestad}.

The fluxes of neutrinos and gamma rays of pionic origin in astrophysical sources are related to the initial accelerated proton spectrum via their final-state multiplicities and average fractional energies $x_{i}$ relative to the proton. Hadronic collisions leads to equal numbers of each type of pion: $p + p \rightarrow \pi^{0}+\pi^{+} + \pi^{-}$. The neutral pion decays into two gamma-rays $\pi^{0} \rightarrow \gamma + \gamma$ and the charged pions into leptons and neutrinos $\pi^{\pm} \rightarrow e + \nu_{\mu} +\nu_{\mu} + \nu_{e}$. The effect of oscillations on the neutrino flux is to equalize the number of neutrinos of each flavor arriving at Earth, giving one neutrino of each flavor per charged pion. Hence we have two-thirds of a charged pion per interacting proton and one neutrino (of each flavor) per charged pion; and one-third of a neutral pion per interacting proton and two photons per neutral pion. Therefore, 

\begin{eqnarray*}
\frac{dN_{\nu}}{dE}\left(E\right) = 1\times \frac{2}{3} \times \frac{1}{x_{\nu}}\times R_{p}\frac{dN_{p}}{dE}\left(\frac{E}{x_{\nu}}\right) \\
\frac{dN_{\gamma}}{dE}\left(E\right) = 2\times \frac{1}{3} \times \frac{1}{x_{\gamma}} \times R_{p}\frac{dN_{p}}{dE}\left(\frac{E}{x_{\gamma}}\right)
\end{eqnarray*}
where $R_{p} = 1-e^{-\tau}$ is the fraction of protons that interact inside the source where $\tau$ is the average number of interactions per proton, analogous to optical depth. We derive the fractional energies from the measured pion inelasticity of approximately 0.2 and from the fact that the final-state particles each take an equal fraction of the pion energy. This gives us $x_{\nu} \simeq 0.25\,x_{\pi}\simeq 0.05$ and $x_{\gamma} \simeq 0.5\,x_{\pi}\simeq 0.1$.

To calculate the initial proton flux we must use the observed Auger cosmic-ray flux, which is what remains of the protons after interacting inside the source: $dN_{cr}/dE = e^{-\tau}dN_{p}/dE$. We calculate the flux assuming a power-law spectrum of the form $dN/dE = N_{0}(E/E_{0})^{\alpha}$, where we can derive the normalization $N_{0}$ from the observational parameters of the Auger experiment by calculating the number of events observed\,\cite{auger}. For $\alpha = -2.0$, we have $N_{\rm events} = {\rm field\,of\,view} \times {\rm time} \times {\rm efficiency} \times N_{0}/E_{\rm thresh}$. This gives us $dN_{cr}/dE = N_{\rm events}\times10^{-13} (E/TeV)^{-2}\,{\rm TeV^{-1}\,cm^{-2}\,s^{-1}}$. Since the gamma-ray flux must be below the limit set by HESS, we can constrain the unknown source parameters $N$ and $\tau$:

\begin{displaymath}
\frac{dN_{\gamma}}{dE}\left(E\right) = \frac{2}{3\,x_{\gamma}}\times\left( e^{\tau}-1\right)\times 10^{-13}\,N\,\left(\frac{E}{x_{\gamma}}\right)^{-2} < 10^{-12} E^{-2}
\end{displaymath}

This gives us the constraint $N(e^{\tau}-1)<150$. The number of events correlating to Cen A depends on the unknown intergalactic magnetic fields. Two correlate within 3\,degrees but as many as ten could be associated with the source. In this case $\tau$ should be ${\mathcal O}$(a few) which is consistent with the physics of the problem previously introduced. Plugging these constants into the expression for the neutrino flux per flavor, we find
\begin{displaymath}
\frac{dN_{\nu}}{dE} \leq 5\times10^{-13} \left( \frac{E}{\rm TeV} \right)^{-2} {\rm TeV^{-1}\,cm^{-2}\,s^{-1}}
\end{displaymath}
With a location at the edge of Auger's sky coverage, a similar calculation for M87 is at present not possible. Clearly, we anticipate the observation with increased statistics of events correlated with M87.

Repeating this calculation for power-law indices between -2.0 and -3.0, we obtain between 0.8 and 0.02 events/year for a generic neutrino detector of effective muon area $\sim1\,{\rm km^{2}}$.

\Section{Diffuse Flux}

\begin{figure}
\epsfig{file=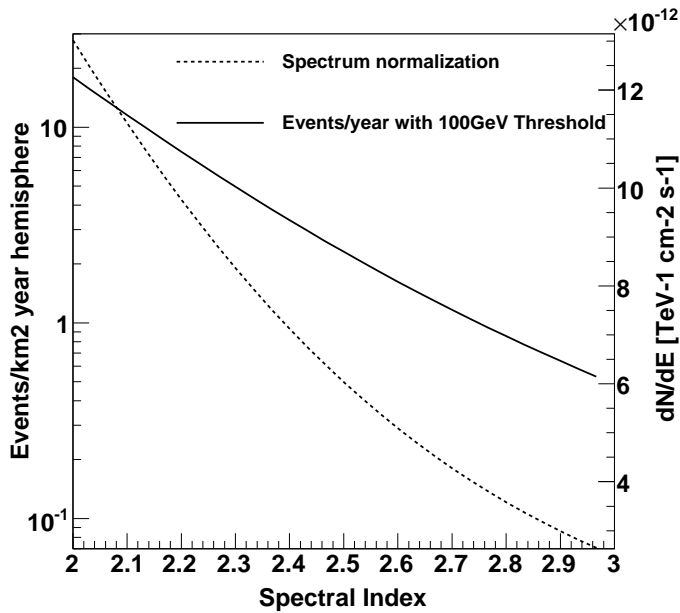, scale=0.45}
\caption{ Total diffuse neutrino flux and event rate from FRI's }
\label{fig:diffevents}
\end{figure}

Having determined the neutrino flux from a UHE cosmic ray source, we can determine the total diffuse flux from all such sources within the horizon. We assume here that the common luminosities of Cen A and M87 are representative of FRI radio galaxies. Given an FRI density of $n \simeq 8\times 10^{4} \,{\rm Gpc^{-3}}$ within a horizon of $R\sim3 \,{\rm Gpc}$\,\cite{fridensity}, the total diffuse flux from all $4\pi \,{\rm sr}$ of the sky is simply the sum of the luminosities of the sources weighted by their distances:
\begin{displaymath}
%\frac{L_{\nu}}{4\pi}\sum\frac{N_{i}}{r_{i}^{2}}=\frac{L_{\nu}}{4\pi}\sum\frac{n4\pi r_{i}^{2}\Delta r}{r_{i}^{2}} = 
\frac{dN_{\nu}}{dE}_{\rm diff} = \sum \frac{L_{\nu}}{4\pi d^{2}} =L_{\nu} \, n\,R = 4\pi d^{2} n R\frac{dN_{\nu}}{dE}_{\rm Cen},
\end{displaymath}
where we perform the sum by assuming the galaxies are uniformly distributed inside the sphere of the horizon. For a Centaurus neutrino flux that goes as $E^{-2}$, this evaluates to
\begin{displaymath}
% = \frac{48.25}{4\pi}\, \frac{dN_{\nu}}{dE}_{Cen}\,{\rm sr^{-1}}
\frac{dN_{\nu}}{dE}_{\rm diff} = 2\times 10^{-9}\,\left(\frac{E}{\rm GeV}\right)^{-2}\,{\rm GeV^{-1}\,cm^{-2}\,s^{-1}\,sr^{-1}},
\end{displaymath}
approximately one-tenth of the Waxman-Bahcall flux\,\cite{WB}. Varying the spectral indices as before, we obtain an event rate per ${\rm km^{2}}$ year from the northern sky of between 19 and 0.5 neutrinos (Figure~\ref{fig:diffevents}). Considering sources out to 3\,Gpc, a redshift of order 0.5, is probably conservative and extending the sources beyond $z\sim1$ may increase the flux by a factor 3. 

\Section{Photohadronic processes}

Often analyses of AGNs assume that any production of secondary particles that serve as a possible signal comes from photohadronic interactions of accelerated charged particles with ambient radiation fields. This means that it is possible to model the emission, in both gamma-rays and neutrinos, of an optically thick ultra-high-energy cosmic ray source\,\cite{ghs}.

Deriving the neutrino and gamma-ray spectra from the proton spectrum proceeds as in the p-p case, with the difference that we must take the different branching ratios of the delta channel $p+\gamma \rightarrow \Delta^{+} \rightarrow p(n)+\pi^{0}(\pi^{+}) $ into account. Due to isospin conservation, the $\pi^{0}$ final state is twice as likely as the $\pi^{+}$ (the opposite of the p-p case), resulting in a relative factor of 4\,\cite{alvarezhalzen}.

One general feature of such a  process is that we expect breaks in the cosmic-ray spectrum corresponding to the onset of energy loss processes, in other words, when the cosmic ray primary energy is large enough to produce pions on the photon field. Following Ref.~\cite{hannestad}, we can then model the gamma-ray and neutrino spectra with a break at the corresponding secondary energy. If we vary the spectral index of the protons, assuming the index hardens by one for energies below the break, we can obtain the neutrino spectrum by fixing the cosmic-ray spectrum above the break at the Auger measurement and fixing the gamma-ray spectrum below the break to the Narrabri measurement\,\cite{cenasydney}. If we take the break energy as variable within a decade of $10^{5}$\,TeV for protons ($\sim 10^{4}$\,TeV for gamma-rays) and allow that the actual number of cosmic-ray events from Cen A is likely different from the Auger experiment's estimate due to the strength of the intergalactic magnetic field being unknown, we can ask which combinations of spectral index, break energy and number of cosmic ray events leads to a continuous spectrum of neutrinos and gamma rays. The final answer is relatively insensitive to the number of cosmic ray events and most sensitive to the spectral index, allowing us to get a spectrum by picking a spectral index and then tuning with the break energy and cosmic ray events. The neutrino event rate per neutrino flavor per ${\rm km^{2}}$ year increases with harder spectra and lower break energies, giving $\sim5$ events/year for a proton spectrum with a a break at $4\times10^{3}$\,TeV and a spectral index of -2.7 above the break, and $\sim0.03$ events/year for a proton spectrum with a break at $5\times10^{5}$\,TeV and a spectral index of -3.0 above the break. We can take these event rates as extreme since the break energies required for continuity are outside the range deemed reasonable. Proton spectral indices harder than -2.7 above the break do not give continuous spectra for break energies above $10^{3}$\,TeV and so were disregarded.

This analysis may be deficient in several regards. On the technical side, it does not account for the cosmic rays being the remnant of the proton spectrum that did not interact in the source since it does not attempt to model the optical depth of the source. Moreover, if the spectral break is due to the onset of pionization, there should be no gamma-rays with energies below $x_{\gamma}E_{break}$, as the source is transparent to the protons at low energies. Thus, in this model the observed TeV gamma ray fluxes from M87 and Cen A cannot be explained. More sensitive observations of Cen A by H.E.S.S. are awaited and confirmation of a TeV flux at the levels assumed in this paper should disfavor the photoproduction model.

In summary, we conclude that FRI modeled on the fluxes measured by the Auger and TeV gamma ray telescopes predict neutrinos at the level of 10\% of the Waxman-Bahcall flux, possibly higher by the factor relating to cosmic evolution mentioned earlier in this paper. There should be no doubt that the predicted flux should be within reach of IceCube\,\cite{icecube} and a future Mediterranean kilometer-scale neutrino telescope, as it is at the level of IceCube's 90\% confidence level sensitivity in a single year\,\cite{footnote}. It follows that detection at the $5\sigma$ level should occur within $\sim5$ years.

\begin{acknowledgments}
The authors would like to thank Steen Hannestad, Julia Becker, and Peter Biermann for valuable discussions. This research was supported in part by the National Science Foundation under Grant No.~OPP-0236449, in part by the U.S.~Department of Energy under Grant No.~DE-FG02-95ER40896, and in part by the University of Wisconsin Research Committee with funds granted by the Wisconsin Alumni Research Foundation.
\end{acknowledgments}


\begin{thebibliography}{[99]}
\bibitem{footnote}
We here used the atmospheric background flux from charm particles predicted by\,\cite{prompt} rather than the larger flux conservatively implemented in the sensitivity calculation presented in the IceCube publication\,\cite{iceperf}. The enhancement in the charm flux relative to the one used here is a direct consequence of the assumption that charm fragmentation functions scale and is therefore unphysical.
\end{thebibliography}
\end{document}